\documentclass{article}

\usepackage{arxiv}
\usepackage{graphicx}
\usepackage[utf8]{inputenc} 
\usepackage[T1]{fontenc}    
\usepackage{hyperref}       
\usepackage{url}            
\usepackage{booktabs}       
\usepackage{amsfonts}       
\usepackage{nicefrac}       
\usepackage{microtype}      
\usepackage{lipsum}
\usepackage{subcaption}
\usepackage{multirow}
\usepackage{bm}

\title{RepBERT: Contextualized Text Embeddings for First-Stage Retrieval}

\author{
  Jingtao Zhan, Jiaxin Mao, Yiqun Liu\thanks{Corresponding author} , Min Zhang, Shaoping Ma \\
  Department of Computer Science and Technology, Institute for Artificial Intelligence \\
  Beijing National Research Center for Information Science and Technology \\
  Tsinghua University, Beijing 100084, China \\
  \texttt{\{jingtaozhan, maojiaxin\}@gmail.com, \{yiqunliu, z-m, msp\}@tsinghua.edu.cn} \\
}

\begin{document}
\maketitle

\begin{abstract}

Although exact term match between queries and documents is the dominant method to perform first-stage retrieval, we propose a different approach, called RepBERT, to represent documents and queries with fixed-length contextualized embeddings. The inner products of query and document embeddings are regarded as relevance scores. On MS MARCO Passage Ranking task, RepBERT achieves state-of-the-art results among all initial retrieval techniques. And its efficiency is comparable to bag-of-words methods.

\end{abstract}


\section{Introduction}

Ranking pipelines are widely used in most search engines. Typically, efficient bag-of-words models are often adopted for initial retrieval, and neural ranking models are utilized for reranking. Although some recent works~\cite{DeepCT,Doc2query,docTTTTTquery} adopt deep language models to improve bag-of-words approaches, they still rely on exact term match signals and can hardly retrieve documents on semantic level. This paper tries to tackle such challenge by directly using deep neural models for first-stage retrieval.

Most neural approaches are time-consuming, especially the well-performing deep language models~\cite{devlin2018bert,T5}. But efficiency is the critical criterion for initial retrieval techniques because each query has millions of candidate documents. To address this, we encode documents into fixed-length embeddings offline and save them to disk to greatly improve online efficiency. During the online retrieval, the model encodes queries and regards inner products between query and document embeddings as relevance scores. The selection of the most relevant documents can be formulated as Maximum Inner Product Search (MIPS), for which many algorithms~\cite{shrivastava2014asymmetric, ram2012maximum, shen2015learning} are proposed and consume sub-linear computational complexity.

BERT~\cite{devlin2018bert} is currently one of the state-of-the-art models in NLP and IR. We adopt it to represent queries and documents. Because our model can be categorized as representation-focused models~\cite{guo2019deep} in IR community, we call the proposed model \textbf{RepBERT}.

This paper adopts the MS MARCO Passage Ranking dataset~\cite{MSMARCO}, which is a benchmark dataset for information retrieval. In the following, we describe in detail how we achieve state-of-the-art results for first-stage retrieval. The code and data are released at \url{https://github.com/jingtaozhan/RepBERT-Index}.

\section{Related Work}
\label{sec:related_work}
Utilizing neural retrievers have been proved to be effective in Open QA tasks~\cite{Realm, karpukhin2020dense} and significantly outperform bag-of-words models, such as BM25~\cite{BM25}. However, bag-of-words models are still the dominant first-stage retrieval approaches in IR community. For example, according to MS MARCO Passage Ranking leaderboard, almost all public methods utilize bag-of-words models for initial retrieval. Such phenomenon may result from some lessons during the early years of neural networks. Prior work~\cite{ARCI} found that encoding text into a fixed-length embedding suffers the risk of losing details and that the interactions between terms are essential for superior ranking performance. But we believe such problem can be solved with powerful language models, such as BERT~\cite{devlin2018bert}.

Despite the lack of neural models for initial retrieval, several works substantially improved bag-of-words models with the help of deep language models. doc2query~\cite{Doc2query} utilizes transformers~\cite{vaswani2017attention} to predict possibly issued queries for a given document and then expands it with those predictions. docTTTTTquery further improves it with the help of T5~\cite{T5} as the expansion model. DeepCT~\cite{DeepCT} uses BERT to compute term weights to replace term frequency field in BM25~\cite{BM25}. 
\section{RepBERT}

\subsection{Model Architectures}
Following BERT's~\cite{devlin2018bert} input style, we apply wordpiece tokenization to the input text, and then add a [CLS] token at the beginning and a [SEP] token at the end:

\begin{equation}
  {\rm Input}(text) = {\rm [CLS]} \quad {\rm Tokenize}(text) \quad {\rm [SEP]}
\end{equation}

Then, we pass the tokens into BERT\footnote{Note that BERT has two segment embeddings, which are added to the embeddings of input tokens in the Embedding Module. In our implementation, we assign segment embeddings numbered $0$ and $1$ to the query tokens and document tokens, respectively.}, which outputs one contextualized vector for each token. The vectors are averaged to produce the contextualized text embedding. In other words, we propose an encoder to represent the input text. Intuitively, representing queries and documents requires similar text understanding ability. Thus, RepBERT shares the weights of query encoder and document encoder. The encoder can be formulated as follows:

\begin{equation}
  {\rm Embed}(text) = {\rm Encoder}(text) = {\rm Average}({\rm BERT}({\rm Input}(text)))
\end{equation}

After acquiring the embeddings of queries and documents, we regard the inner products of them as relevance scores. Such simple design is mainly based on efficiency considerations. It can be formulated as follows:

\begin{equation}
  {\rm Rel}(query, doc) = {\rm Embed}(query)^\top \cdot {\rm Embed}(doc)
\end{equation}

\subsection{Training}

\textbf{Loss Function} The goal of training is to make the embedding inner products of relevant pairs of queries and documents larger than those of irrelevant pairs. Let $(q, d_1^+, ..., d_m^+, d_{m+1}^-, ..., d_n^-)$ be one instance of the input training batch. The instance contains one query $q$, $m$ relevant (positive) documents and $n-m$ irrelevant (negative) documents. We adopt MultiLabelMarginLoss~\cite{paszke2017automatic} as the loss function:

\begin{equation}
  \label{eq:loss_func}
  \mathcal{L}(q, d_1^+, ..., d_m^+, d_{m+1}^-, ..., d_n^-) = \frac{1}{n} \cdot \sum_{1 \leq i \leq m, m < j \leq n} {\rm max}(0, 1-({\rm Rel}(q, d_i^+)-{\rm Rel}(q, d_j^-)))
\end{equation}

\textbf{In-batch Negatives}: During training, it is computationally expensive to sample many negative documents for each query. The trick of in-batch negatives is to utilize the documents from other query-document pairs in the same mini-batch as negative examples. For instance, there are $B$ query-document pairs in the mini-batch. Thus, most of the time, each query has $1$ positive example and $B-1$ negative examples. In rare cases, for a given query, some documents from other query-document pairs (the usual $B-1$ negatives) may be relevant and thus are regarded as positive in Equation~\ref{eq:loss_func}. Such trick has been used in prior works~\cite{karpukhin2020dense, gillick2019learning} for training a siamese neural network.

\section{Experiment}

\subsection{Dataset}
MS MARCO Passage Ranking Dataset~\cite{MSMARCO} (MS MARCO) is a benchmark English dataset for ad-hoc retrieval. It has approximately 8.8 million passages, 0.5 million queries for training, 6.9 thousand queries for development. A blind, held-out evaluation set with about 6.8 thousand queries is also available and the result is provided by the organizers upon submission to the online leaderboard. In order to maintain consistent terminology throughout this paper, we refer to these basic units of retrieval as "documents".

\subsection{Baselines}

We compare with four initial retrieval techniques public on MS MARCO leaderboard, which are BM25(Anserini)~\cite{yang2018anserini}, doc2query~\cite{Doc2query}, DeepCT~\cite{DeepCT}, and docTTTTTquery~\cite{docTTTTTquery}. The last three methods use deep language models to improve BM25 and are very competitive. They are briefly introduced in Section~\ref{sec:related_work}.

We also show performances of two-stage retrieval techniques. BiLSTM + Co-Attention + self attention based document scorer~\cite{Alaparthi2019MicrosoftAC} is the best non-ensemble, non-BERT method from the leaderboard with an associated paper. It uses BM25 for initial retrieval and deep attention networks for reranking. Another technique is proposed by Nogueira et al.~\cite{nogueira2019passage}, which uses BM25 for initial retrieval and BERT Large for reranking.

\subsection{First-Stage Retrieval}

This section compares RepBERT with other retrieval techniques based on the performance of first-stage retrieval.

\begin{table}
    \caption{Performances of first-stage retrieval and two-stage retrieval models on MS MARCO Passage Ranking dataset}
    \centering
    \begin{tabular}{llll|l}
    \toprule
    & \multicolumn{2}{c}{MRR@10} & R@1000 & Latency  \\
    & Dev & Test & Dev & (ms/query) \\
    \midrule
    BM25(Anserini)~\cite{Doc2query} & 0.184 & 0.186 & 0.853 & 50 \\
    doc2query~\cite{Doc2query} & 0.215 & 0.218 & 0.893 & 90 \\
    DeepCT~\cite{DeepCT} & 0.243 & 0.239 & 0.913 & 55 \\
    docTTTTTquery~\cite{docTTTTTquery} & 0.277 & 0.272 & \textbf{0.947} & 64 \\
    Ours (RepBERT) & \textbf{0.304} & \textbf{0.294} & 0.943 & 80 \\
    \midrule
    Best non-ensemble, non-BERT~\cite{Alaparthi2019MicrosoftAC} & 0.298 & 0.291 & 0.814 & - \\
    BM25 + BERT Large~\cite{nogueira2019passage} & 0.365 & 0.358 & 0.814 & 3,400 \\
    \bottomrule
    \end{tabular}
    \label{tab:FirstStage_MSMARCO_results}
\end{table}

\subsubsection{Settings}

We adopt the "Train Triples" data provided by MS MARCO~\cite{MSMARCO} for training. Due to the limitation of computational resources, we adopt the BERT base model in our experiment, which consists of $12$ encoder layers with vector dimension of $768$. The maximum query length and document length are set to $20$ and $256$ tokens, respectively. We fine-tune the model using one Titan XP GPU with a batch size of $26$ and gradient accumulation steps of $2$ for $350 \rm k$ steps, which corresponds to training on $\rm 18.2M$ ($\rm 350k \times 26 \times 2$) query-document pairs. We could not observe any improvement based on a small dev set when training for another $\rm 100k$ steps.

Our implementation is based on a public transformer library~\cite{Wolf2019HuggingFacesTS}. We follow the hyper parameter settings in Rodrigo et al.~\cite{nogueira2019passage}. Specifically, we use ADAM~\cite{Adam} with the initial learning rate set to $3 \times 10^{-6}$, $\beta_1 = 0.9$, $\beta_2 = 0.999$, L2 weight decay of $0.01$, learning rate warmup over the first $10,000$ steps, and linear decay of the learning rate. We use a dropout probability of $0.1$ on all layers.

The latency of different models is also provided. The latency of baselines are copied from prior works~\cite{Doc2query, docTTTTTquery}. As for our models, because the document embeddings consume $26 \rm GB$ and thus are impossible to load into a single $12 \rm GB$ GPU, we utilize $3$ Titan XP and $2$ GeForce GTX 1080ti to retrieve top-1000 documents for each query. We report the average latency to retrieve queries in the dev set. The efficiency can be further improved using more advanced GPUs or TPUs.

\subsubsection{Discussion}

The results are shown in Table~\ref{tab:FirstStage_MSMARCO_results}.

RepBERT can represent text to retrieve documents on semantic level with high accuracy. Considering the MRR@10 metric, our model substantially outperforms other first-stage retrieval techniques. Particularly, it is better than the best non-ensemble, non-BERT two-stage retrieval method.

RepBERT can achieve high recall and thus its ranking results can be used for subsequent reranking models. Considering the Recall@1000 metric, our model is very near the best result achieved by docTTTTTquery~\cite{docTTTTTquery}, which utilizes more powerful T5~\cite{T5} language model. It significantly outperforms other baselines. We believe using more advanced language models to represent text can further improve RepBERT, just as how docTTTTTquery improves doc2query. 

In terms of efficiency, RepBERT is comparable to bag-of-words models. It shows that it is practical to represent documents offline and compute inner products online for first-stage retrieval. Note that in our current retrieval implementation, we have not adopted optimized MIPS algorithms~\cite{shrivastava2014asymmetric, ram2012maximum, shen2015learning} and simply compute relevance scores between the given query and each document. We plan to investigate them in the future.

In summary, we propose a method to represent text with fixed-length embeddings and efficiently retrieve documents with high accuracy and recall. The model outperforms the original or the improved bag-of-words models, which highlights the possibility to replace them for initial retrieval.

\subsection{Rerank based on RepBERT}

This section investigates the performance of a reranking model when using RepBERT as the first-stage retriever.

\begin{figure}[t]
    \caption{At different depths, the recall of the first-stage retrieval method and the reranking accuracy of BERT Large. Dataset: MS MARCO dev.}
    \hspace*{\fill}%
    \subcaptionbox{Recall at different depths\label{fig:recall}}
    {\includegraphics[width=.48\linewidth]{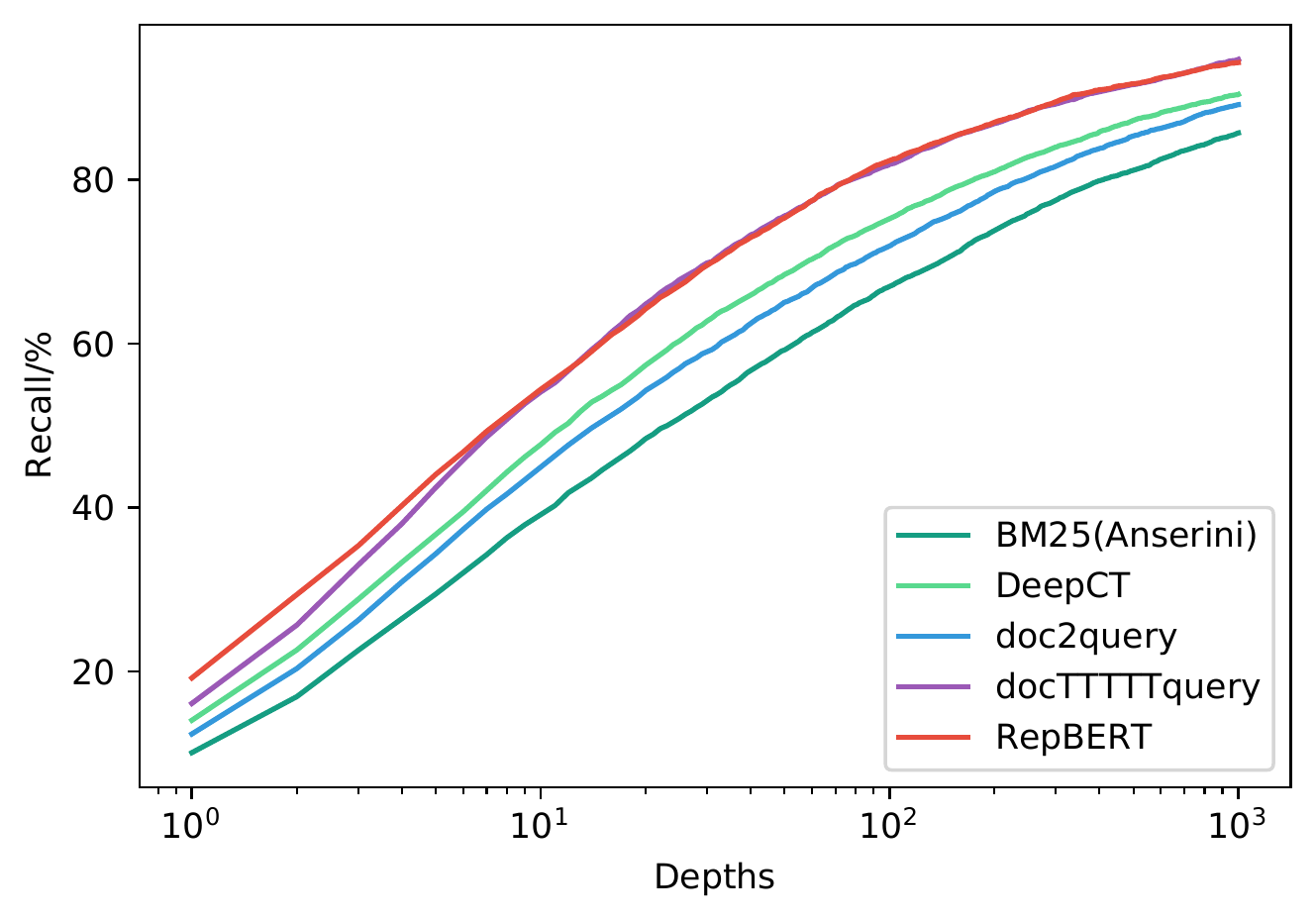}}
    \hfill%
    \subcaptionbox{Reranking Performance at different depths\label{fig:performance}}
    {\includegraphics[width=.48\linewidth]{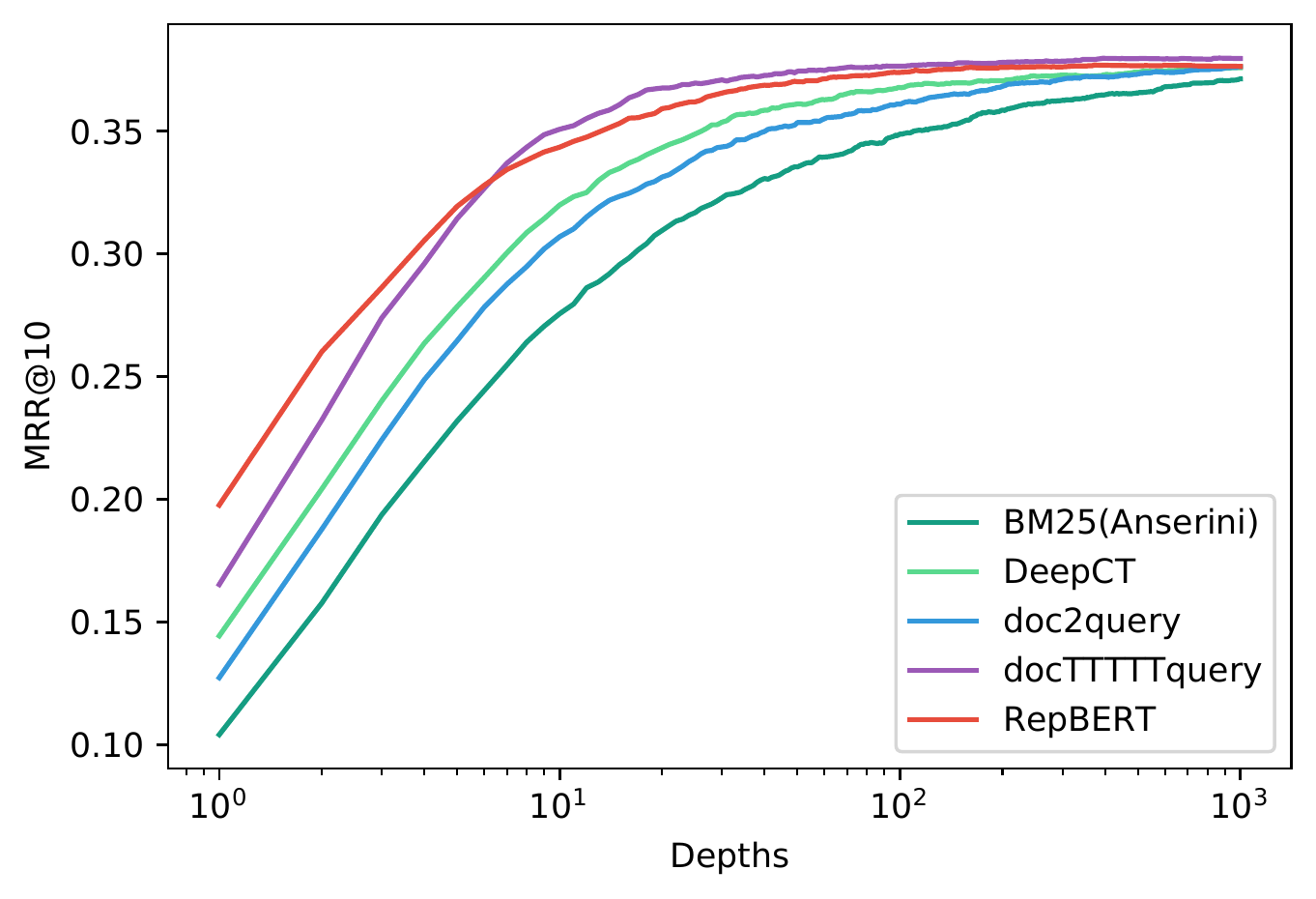}}%
    \hspace*{\fill}%
    \label{fig:recall_performance}
\end{figure}

\begin{table}
    \caption{Reranking accuracy (MRR@10) of BERT Large~\cite{nogueira2019passage} using different first-stage retrieval techniques at different reranking depths. Dataset: MS MARCO dev. The improvement is relative to the reranking performance using BM25(Anserini) index.}
    \centering
    \begin{tabular}{l|cllll}
    \toprule
    Depths & BM25(Anserini) & doc2query & DeepCT &  docTTTTTquery & RepBERT \\
    \midrule
    5 &  0.232 & 0.265 ($+$14\%) & 0.279 ($+$20\%) & 0.314 ($+$36\%) & \textbf{0.319 (}$\bm{+}$\textbf{38\%)} \\
    10 &  0.276 & 0.307 ($+$11\%) & 0.320 ($+$16\%) & \textbf{0.351 (}$\bm{+}$\textbf{27\%)} & 0.344 ($+$25\%) \\
    50 & 0.336 & 0.354 ($+$5\%) & 0.361 ($+$8\%) & \textbf{0.375 (}$\bm{+}$\textbf{12\%)} & 0.370 ($+$10\%) \\
    500 & 0.366 & 0.373 ($+$2\%) & 0.374 ($+$2\%) & \textbf{0.380 (}$\bm{+}$\textbf{4\%)} & 0.377 ($+$3\%) \\
    1000 &  0.371 & 0.376 ($+$1\%) & 0.376 ($+$1\%) & \textbf{0.380 (}$\bm{+}$\textbf{2\%)} & 0.376 ($+$1\%) \\
    \bottomrule
    \end{tabular}
    \label{tab:Rerank_MSMARCO_results}
\end{table}

\subsubsection{Settings}

Intuitively, the recall rate is an important factor for reranking performance. Thus, we compute it for different first-stage retrieval techniques at different depths. The results are shown in Figure~\ref{fig:recall}.

Following prior works~\cite{DeepCT, Doc2query}, we directly utilize the public BERT Large model~\cite{nogueira2019passage} finetuned on MS MARCO to rerank the documents retrieved by different models, except doc2query~\cite{Doc2query} which already made the reranking run file public. The overall performances on dev set at different depths are shown in Table~\ref{tab:Rerank_MSMARCO_results} and Figure~\ref{fig:performance}.

\subsubsection{Discussion}

According to Figure~\ref{fig:recall}, our proposed RepBERT can achieve the best recall rates at small depths, partly due to the highest retrieval accuracy of our model. At large depths, RepBERT and docTTTTTquery are both the best-performing models. Thus, RepBERT's reranking performances should be the best at all depths. 

According to Table~\ref{tab:Rerank_MSMARCO_results} and Figure~\ref{fig:performance}, using RepBERT can achieve the best results at small depths. At large depths, such as 50, though docTTTTTquery's performance is the best, using RepBERT can significantly outperform other baselines. At larger depths, such as 500 or 1000, the performance gap between models becomes smaller. 

However, there is some inconsistency between the recall and the reranking performances. Although at large depths, RepBERT is still as good as docTTTTTquery in terms of recall, its reranking performances are worse than docTTTTTquery. We believe such inconsistency is due to the mismatch between training and testing data distribution for reranking model. It is elaborated in the next section.

\subsubsection{Mismatch}

\begin{figure}[t]
    \centering
    \caption{For a certain depth, the average proportion of retrieved documents that are also in the official top-1000 candidates provided by MS MARCO. Dataset: MS MARCO dev.}
    \includegraphics[width=0.48\linewidth, keepaspectratio=True]{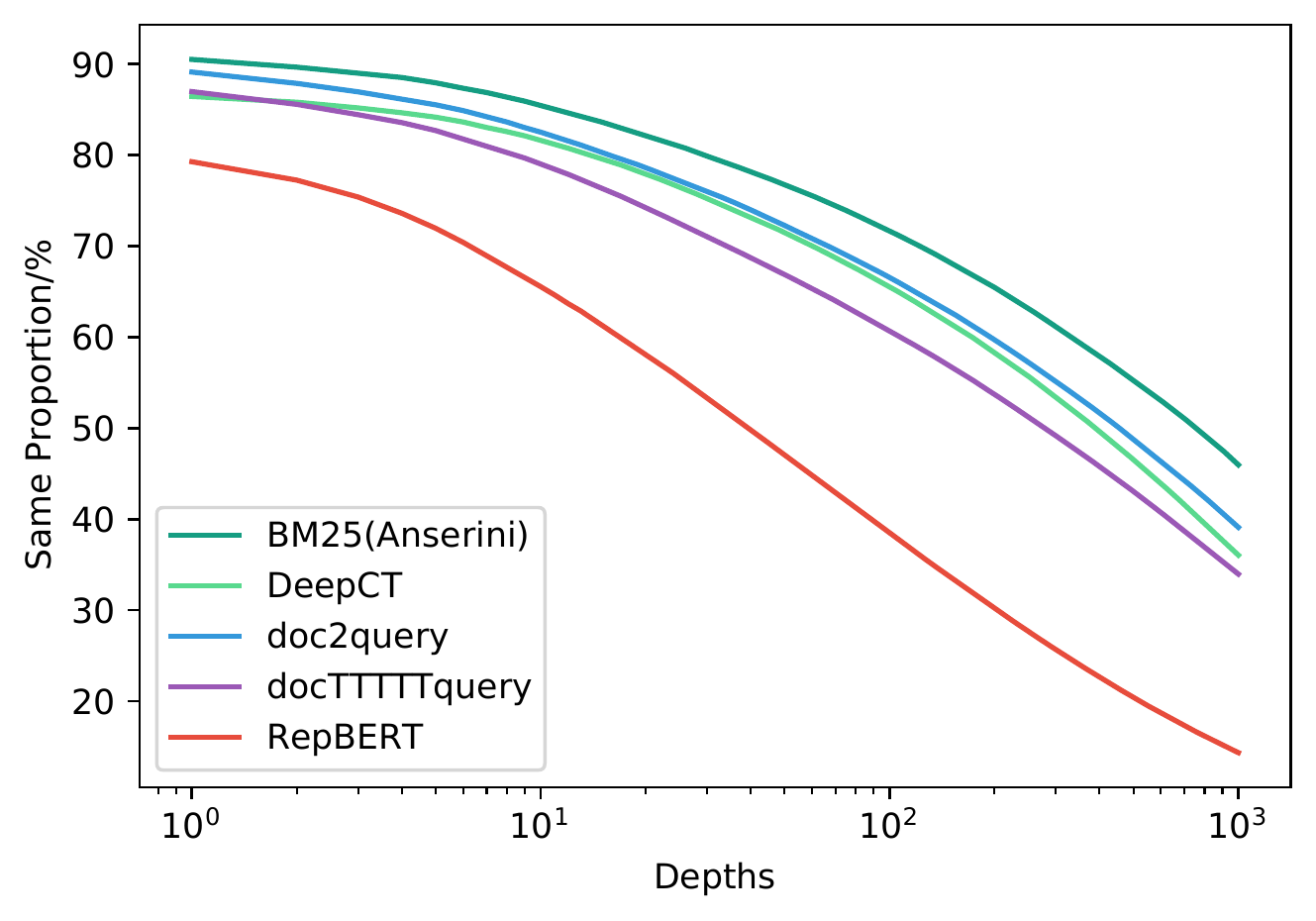}
    \label{fig:depths_consistency}
\end{figure}

In the following, we present our speculation that the mismatch leads to performance loss. The reranking model~\cite{nogueira2019passage} used in prior works~\cite{DeepCT,Doc2query} and ours is trained based on the "Train Triples" data provided by MS MARCO. It was generated by pairing positive documents in the qrel file with the negative documents in the top-1000 file retrieved by the official BM25~\footnote{\url{https://github.com/microsoft/TREC-2019-Deep-Learning/issues/15}}. However, during testing, the model is used to rerank the documents retrieved by another method, such as RepBERT. It can cause severe mismatch of input data distribution if the retrievers used during training and testing are very different.

Before further elaboration, we introduce several denotations. We use $F$, $q$, and $n$ ($n \leq 1000$) to denote a retrieval technique, a query, and a depth, respectively. Specifically, the new technique used in testing is denoted as $f$ and the official retriever used in training is denoted as $BM25$. We use $\mathcal{D}_{F, q, n}$ to denote the top-n documents retrieved by $F$ for a given query $q$. Note that MS MARCO's "Train Triples" data is generated using $\mathcal{D}_{BM25, q, 1000}$. 

We use a simple method to quantify such mismatch based on an intuitive thought. If $\mathcal{D}_{f, q, n} \sqsubseteq \mathcal{D}_{BM25, q, 1000}$, there is no mismatch for query $q$ at depth $n$. But if $\mathcal{D}_{f, q, n} \cap \mathcal{D}_{BM25, q, 1000} = \emptyset$ , the mismatch for query $q$ at depth $n$ is the biggest. Therefore, we define the consistency factor of $f$ at depth $n$, called $C_{f, n}$, as the average proportion of documents in $\mathcal{D}_{f, q, n}$ that are also in $\mathcal{D}_{BM25, q, 1000}$. Thus, $C_{f, n} \in [0, 1]$, where $1$ represents no mismatch and $0$ represents the biggest mismatch. It can be formulated as follows ($|X|$ is the cardinality of set $X$.):

\begin{equation}
    \label{eq:consistency}
    C_{f, n} = \frac{1}{|\{q\}|} \sum_{q} \frac{| \mathcal{D}_{f, q, n}  \cap \mathcal{D}_{BM25, q, 1000} |}{| \mathcal{D}_{f, q, n}|} 
\end{equation}

We compute the $C_{f, n}$ for different first-stage retrieval techniques, and the results are shown in Figure~\ref{fig:depths_consistency}. The consistency factor of RepBERT is significantly lower than other methods, especially at large depths. It means that a major proportion of documents retrieved by RepBERT are not considered as candidates by the official BM25. Such results agree with the design of different techniques. The prior works, though also using deep language models, still relies on exact match signals to retrieve, while our proposed model utilizes semantic match signals. The distribution of retrieved documents between RepBERT and BM25 is very different. Thus the model trained to rerank documents retrieved by BM25 may not work well when reranking documents retrieved by RepBERT. We believe training BERT Large with negatives sampled from top-1000 documents retrieved by RepBERT can solve this issue.

\subsection{Combination of Semantic Match and Exact Match}

As introduced in the previous section, RepBERT utilizes semantic match signals, which is very different from BM25 and its improved versions using exact match signals. Thus, it is a straightforward idea to investigate whether the combination of the two signals can achieve better retrieval performance.

\subsubsection{Method}

Before we present a simple method to combine two retrieval techniques, we introduce several denotations. 
We use $F$ to denote a retrieval technique, and $d_{F, q, i}$ to denote the $i_{th}$ document retrieved by $F$ for a given query $q$. Thus, the top-n documents retrieved by $F$ for a given query $q$, denoted as $\mathcal{D}_{F, q, n}$, are equal to $[d_{F, q, 1}, d_{F, q, 2}, ..., d_{F, q, n}]$. 

Our method is as follows. We use $f_a$ and $f_b$ to refer to the two techniques that will be combined. First, we merge the two retrieval document lists, namely $\mathcal{D}_{f_a, q, 1000}$ and $\mathcal{D}_{f_b, q, 1000}$, in an alternating fashion to acquire a preliminary ranking list. For example, if $\mathcal{D}_{f_a, q, 3} = [a, c, d]$ and $\mathcal{D}_{f_b, q, 3} = [b, a ,c]$, then the merged list is $[a,b,c,a,d,c]$. Such operation usually results in duplicated documents at different ranking positions. Thus, we filter out the documents that also appear at lower ranking positions. In our example, the filtered list is $[a,b,c,d]$. Finally, we truncate the filtered list to contain only the first 1000 documents. The whole process can be formulated as follows:

\begin{equation}
    \label{eq:combine}
    \mathcal{D}_{f_a+f_b, q, 1000} = {\rm Truncate}({\rm Filter}([d_{f_a, q, 1}, d_{f_b, q, 1}, d_{f_a, q, 2}, d_{f_b, q, 2}, ..., d_{f_a, q, 1000}, d_{f_b, q, 1000},]))
\end{equation}

We present the retrieval accuracy and recall of different combinations in Table~\ref{tab:combine_accuracy} and ~\ref{tab:combine_recall}, respectively. The cell in the $a_{th}$ row and $b_{th}$ column shows the performance of $f_a+f_b$ and the improvement compared with $f_a$. Note that in our method, $f_a+f_b$ and $f_b+f_a$ are different combinations, which is clearly reflected using MRR@10 metric.

It is worth pointing out that there is much room for improvement. For example, without truncation in Equation~\ref{eq:combine}, Recall@2000 is \textbf{0.980} for RepBERT+docTTTTTquery, compared with 0.967 after truncation. There may be other methods to achieve better combining performance.

\begin{table}
    \caption{The retrieval accuracy (MRR@10) of different technique combinations. Dataset: MS MARCO dev. The cell in the $a_{th}$ row and $b_{th}$ column shows the ranking accuracy of the combination and the improvement compared with the model corresponding to the $a_{th}$ row.}
    \centering
    \begin{tabular}{l|lllll}
    \toprule
    MRR@10 & +BM25(Anserini) & +doc2query & +DeepCT & +docTTTTTquery & +RepBERT \\
    \midrule
    BM25(Anserini) & 0.187 & 0.203 ($+8\%$) & 0.213 ($+13\%$) & 0.227 ($+21\%$) & \textbf{0.245 (}$\bm{+}$\textbf{31\%)} \\
    doc2query & 0.217 ($-2\%$) & 0.222 & 0.236 ($+6\%$) & 0.247 ($+12\%$) & \textbf{0.263 (}$\bm{+}$\textbf{19\%)} \\
    DeepCT & 0.236 ($-$3\%) & 0.246 ($+$1\%) & 0.243 & 0.263 ($+$8\%) & \textbf{0.276 (}$\bm{+}$\textbf{14\%)} \\
    docTTTTTquery &  0.263 ($-$5\%) & 0.270 ($-$3\%) & 0.275 ($-$1\%) & 0.277 & \textbf{0.298 (}$\bm{+}$\textbf{8\%)} \\
    RepBERT &  0.296 ($-$2\%) & 0.302 ($-$1\%) & 0.306 ($+$1\%) & \textbf{0.315 (}$\bm{+}$\textbf{4\%)} & 0.304 \\
    \bottomrule
    \end{tabular}
    \label{tab:combine_accuracy}
\end{table}

\begin{table}
    \caption{The retrieval recall of different technique combinations. Dataset: MS MARCO dev. The cell in the $a_{th}$ row and $b_{th}$ column shows the recall of the combination and the improvement compared with the model corresponding to the $a_{th}$ row.}
    \centering
    \begin{tabular}{l|lllll}
    \toprule
    Recall@1000 & +BM25(Anserini) & +doc2query & +DeepCT & +docTTTTTquery & +RepBERT \\
    \midrule
    BM25(Anserini) & 0.857 & 0.888 ($+$4\%) & 0.909 ($+$6\%) & 0.937 ($+$9\%) & \textbf{0.957 (}$\bm{+}$\textbf{12\%)} \\
    doc2query & 0.888 ($-$0\%) & 0.892 & 0.919 ($+$3\%) & 0.941 ($+$6\%) & \textbf{0.961 (}$\bm{+}$\textbf{8\%)} \\
    DeepCT & 0.909 ($+$1\%) & 0.919 ($+$2\%) & 0.904 & 0.949 ($+$5\%) & \textbf{0.957 (}$\bm{+}$\textbf{6\%)} \\
    docTTTTTquery &  0.937 ($-$1\%) & 0.942 ($-$1\%) & 0.949 ($+$0\%) & 0.947 & \textbf{0.967 (}$\bm{+}$\textbf{2\%)} \\
    RepBERT &  0.957 ($+$1\%) & 0.961 ($+$2\%) & 0.957 ($+$1\%) & \textbf{0.967 (}$\bm{+}$\textbf{3\%)} & 0.943 \\
    \bottomrule
    \end{tabular}
    \label{tab:combine_recall}
\end{table}

\subsubsection{Discussion}

As shown in Table~\ref{tab:combine_accuracy} and ~\ref{tab:combine_recall}, BM25 and the improved versions achieve the best ranking accuracy and recall when combined with RepBERT. Especially in terms of recall, although docTTTTTquery is as good as (or slightly better than) RepBERT according to Table~\ref{tab:FirstStage_MSMARCO_results}, RepBERT can better boost the recall of other baselines. We believe it is because RepBERT can better complement the semantic matching ability lacked by these baselines. 

For RepBERT, combinations with exact match retriever can improve its recall and may also improve its ranking accuracy. According to Table~\ref{tab:FirstStage_MSMARCO_results}, ~\ref{tab:combine_accuracy}, and ~\ref{tab:combine_recall}, RepBERT+docTTTTTquery is the best first-stage retriever in this paper. The results suggest that exact match signals are also helpful for semantic matching retrievers. 
\section{Conclusion}

This paper proposes RepBERT to represent text with contextualized embeddings for first-stage retrieval. It achieves state-of-the-art initial retrieval performance on MS MARCO Passage Ranking dataset. We highlight the possibility to use representation-focused neural models to replace the widely-adopted bag-of-words models in first-stage retrieval. In the future, we plan to test model's generalization ability on different datasets and investigate its performance in retrieving long text.

\bibliographystyle{unsrt}  
\bibliography{references}  

\begin{thebibliography}{10}

\bibitem{DeepCT}
Zhuyun Dai and Jamie Callan.
\newblock Context-aware sentence/passage term importance estimation for first
  stage retrieval.
\newblock {\em arXiv preprint arXiv:1910.10687}, 2019.

\bibitem{Doc2query}
Rodrigo Nogueira, Wei Yang, Jimmy Lin, and Kyunghyun Cho.
\newblock Document expansion by query prediction.
\newblock {\em arXiv preprint arXiv:1904.08375}, 2019.

\bibitem{docTTTTTquery}
Rodrigo Nogueira, Jimmy Lin, and AI~Epistemic.
\newblock From doc2query to doctttttquery.
\newblock {\em Online preprint}, 2019.

\bibitem{devlin2018bert}
Jacob Devlin, Ming-Wei Chang, Kenton Lee, and Kristina Toutanova.
\newblock Bert: Pre-training of deep bidirectional transformers for language
  understanding.
\newblock {\em arXiv preprint arXiv:1810.04805}, 2018.

\bibitem{T5}
Colin Raffel, Noam Shazeer, Adam Roberts, Katherine Lee, Sharan Narang, Michael
  Matena, Yanqi Zhou, Wei Li, and Peter~J Liu.
\newblock Exploring the limits of transfer learning with a unified text-to-text
  transformer.
\newblock {\em arXiv preprint arXiv:1910.10683}, 2019.

\bibitem{shrivastava2014asymmetric}
Anshumali Shrivastava and Ping Li.
\newblock Asymmetric lsh (alsh) for sublinear time maximum inner product search
  (mips).
\newblock In {\em Advances in neural information processing systems}, pages
  2321--2329, 2014.

\bibitem{ram2012maximum}
Parikshit Ram and Alexander~G Gray.
\newblock Maximum inner-product search using cone trees.
\newblock In {\em Proceedings of the 18th ACM SIGKDD international conference
  on Knowledge discovery and data mining}, pages 931--939, 2012.

\bibitem{shen2015learning}
Fumin Shen, Wei Liu, Shaoting Zhang, Yang Yang, and Heng Tao~Shen.
\newblock Learning binary codes for maximum inner product search.
\newblock In {\em Proceedings of the IEEE International Conference on Computer
  Vision}, pages 4148--4156, 2015.

\bibitem{guo2019deep}
Jiafeng Guo, Yixing Fan, Liang Pang, Liu Yang, Qingyao Ai, Hamed Zamani, Chen
  Wu, W~Bruce Croft, and Xueqi Cheng.
\newblock A deep look into neural ranking models for information retrieval.
\newblock {\em Information Processing \& Management}, page 102067, 2019.

\bibitem{MSMARCO}
Payal Bajaj, Daniel Campos, Nick Craswell, Li~Deng, Jianfeng Gao, Xiaodong Liu,
  Rangan Majumder, Andrew McNamara, Bhaskar Mitra, Tri Nguyen, et~al.
\newblock Ms marco: A human generated machine reading comprehension dataset.
\newblock {\em arXiv preprint arXiv:1611.09268}, 2016.

\bibitem{Realm}
Kelvin Guu, Kenton Lee, Zora Tung, Panupong Pasupat, and Ming-Wei Chang.
\newblock Realm: Retrieval-augmented language model pre-training.
\newblock {\em arXiv preprint arXiv:2002.08909}, 2020.

\bibitem{karpukhin2020dense}
Vladimir Karpukhin, Barlas O{\u{g}}uz, Sewon Min, Ledell Wu, Sergey Edunov,
  Danqi Chen, and Wen-tau Yih.
\newblock Dense passage retrieval for open-domain question answering.
\newblock {\em arXiv preprint arXiv:2004.04906}, 2020.

\bibitem{BM25}
Stephen Robertson and Hugo Zaragoza.
\newblock {\em The probabilistic relevance framework: BM25 and beyond}.
\newblock Now Publishers Inc, 2009.

\bibitem{ARCI}
Baotian Hu, Zhengdong Lu, Hang Li, and Qingcai Chen.
\newblock Convolutional neural network architectures for matching natural
  language sentences.
\newblock In {\em Advances in neural information processing systems}, pages
  2042--2050, 2014.

\bibitem{vaswani2017attention}
Ashish Vaswani, Noam Shazeer, Niki Parmar, Jakob Uszkoreit, Llion Jones,
  Aidan~N Gomez, {\L}ukasz Kaiser, and Illia Polosukhin.
\newblock Attention is all you need.
\newblock In {\em Advances in neural information processing systems}, pages
  5998--6008, 2017.

\bibitem{paszke2017automatic}
Adam Paszke, Sam Gross, Soumith Chintala, Gregory Chanan, Edward Yang, Zachary
  DeVito, Zeming Lin, Alban Desmaison, Luca Antiga, and Adam Lerer.
\newblock Automatic differentiation in pytorch.
\newblock 2017.

\bibitem{gillick2019learning}
Daniel Gillick, Sayali Kulkarni, Larry Lansing, Alessandro Presta, Jason
  Baldridge, Eugene Ie, and Diego Garcia-Olano.
\newblock Learning dense representations for entity retrieval.
\newblock {\em arXiv preprint arXiv:1909.10506}, 2019.

\bibitem{yang2018anserini}
Peilin Yang, Hui Fang, and Jimmy Lin.
\newblock Anserini: Reproducible ranking baselines using lucene.
\newblock {\em Journal of Data and Information Quality (JDIQ)}, 10(4):1--20,
  2018.

\bibitem{Alaparthi2019MicrosoftAC}
Chaitanya~Sai Alaparthi.
\newblock Microsoft ai challenge india 2018: Learning to rank passages for web
  question answering with deep attention networks.
\newblock {\em ArXiv}, abs/1906.06056, 2019.

\bibitem{nogueira2019passage}
Rodrigo Nogueira and Kyunghyun Cho.
\newblock Passage re-ranking with bert.
\newblock {\em arXiv preprint arXiv:1901.04085}, 2019.

\bibitem{Wolf2019HuggingFacesTS}
Thomas Wolf, Lysandre Debut, Victor Sanh, Julien Chaumond, Clement Delangue,
  Anthony Moi, Pierric Cistac, Tim Rault, R'emi Louf, Morgan Funtowicz, and
  Jamie Brew.
\newblock Huggingface's transformers: State-of-the-art natural language
  processing.
\newblock {\em ArXiv}, abs/1910.03771, 2019.

\bibitem{Adam}
Diederik~P Kingma and Jimmy Ba.
\newblock Adam: A method for stochastic optimization.
\newblock {\em arXiv preprint arXiv:1412.6980}, 2014.

\end{thebibliography}

\end{document}